\documentclass[a4paper,11pt]{article}
\usepackage{jheppub}
\usepackage{graphicx}
\usepackage{amsmath}
\usepackage{amsfonts}
\usepackage{amssymb}
\usepackage{slashed}
\usepackage[colorlinks=true,linkcolor=blue]{hyperref}
\usepackage{bm}
\usepackage{empheq}
\usepackage[super]{nth}

\def\argphi{\,\mathrm{Arg}\,\varphi}
\def\tA{\theta_{\mathrm{a}}}
\def\rhodm{\rho_{\mathrm{dm}}}
\def\mrsq{(m_a/m)^2_{\mathrm{max}}}

\def\tstart{t_{\mathrm{start}}}

\def\kmax{k_{\mathrm{max}}}

\def\OO{\mathcal{O}}
\def\LL{\mathcal{L}}
\def\be{\begin{equation}}
\def\ee{\end{equation}}
\def\bea{\begin{eqnarray}}
\def\eea{\end{eqnarray}}

\def\Eq#1{Eq.~\eqref{#1}}

\def\nax{n_{\mathrm{ax}}}
\def\nmiss{n_{\mathrm{misalign}}}

\title{The dark-matter axion mass}
\author{Vincent B.\ Klaer, Guy D.\ Moore}

\affiliation{Institut f\"ur Kernphysik, Technische Universit\"at Darmstadt\\
Schlossgartenstra{\ss}e 2, D-64289 Darmstadt, Germany}
\emailAdd{vklaer@theorie.ikp.physik.tu-darmstadt.de,guy.moore@physik.tu-darmstadt.de}

\abstract{
We evaluate the efficiency of axion production from spatially random
initial conditions in the axion field, so a network of axionic strings
is present.  For the first time, we perform numerical simulations
which fully account for the large short-distance contributions to the
axionic string tension, and the resulting dense network of
high-tension axionic strings.  We find nevertheless that the total axion
production is somewhat \textsl{less} efficient than in the
angle-averaged misalignment case.  Combining our results with a recent
determination of the hot QCD topological susceptibility
\cite{Borsanyi:2016ksw}, we find that if the axion makes up all of the
dark matter, then the axion mass is
$m_a = 26.2\pm 3.4\:\mu\mathrm{eV}$.
}

\begin{document}
\maketitle
\section{Introduction}
\label{sec:intro}

The QCD axion \cite{Weinberg:1977ma,Wilczek:1977pj}
is a hypothetical particle, predicted in models which solve the
QCD theta problem \cite{tHooft:1976,Jackiw:1976pf,Callan:1979bg}
via the Peccei-Quinn mechanism \cite{Peccei:1977hh,Peccei:1977ur}.
In the simplest models \cite{Kim:1979if,Shifman:1979if}
it is the angular mode of a complex scalar,
$\sqrt{2}\,\varphi = f e^{i\tA}$ or $\tA=\argphi$, which would be the
Goldstone boson of a spontaneously broken U(1) symmetry, except that
the symmetry is also anomalous (explicitly broken by QCD).  Therefore
QCD effects induce a small, temperature-dependent ``tilt'' in the
potential, and therefore make the angular fluctuations massive,
so the axion mass is $m_a \neq 0$.   The axion is a dark matter
candidate \cite{Preskill:1982cy,Abbott:1982af,Dine:1982ah}
because early Universe dynamics generically generate large coherent
oscillations in the axion field -- essentially a Bose-Einstein
condensate of axions at or near rest -- which act as a pressureless
fluid on scales longer than a few meters.

In this paper we will predict the axion's mass, given the following
hypotheses:
\begin{enumerate}
\item
  the axion exists;
\item
  PQ symmetry is restored either during or at some point after
  inflation, so that the axion field starts out ``random,'' meaning
  that its value at points out of post-inflationary causal contact are
  uncorrelated \cite{Visinelli:2009zm,Visinelli:2014twa};
\item
  The cosmological epoch where axions are produced -- roughly,
  temperatures around $1\:\mathrm{GeV}$ -- follows standard
  FRW behavior with the expected standard-model matter content;
\item
  the axion makes up $100\%$ of the dark matter, so its current energy
  density is set by measurements of $\Omega_{\mathrm{dm}} h^2$;
  $\rhodm/s = 0.39 \: \mathrm{eV}$ with $s$ the
  entropy density \cite{Ade:2015xua}.
\end{enumerate}
These assumptions give rise to a rich dynamics, with a network of
axionic cosmic strings \cite{Davis:1986xc}
which collapses once the axion mass becomes large in units of the
system age, $m_a t \gg 1$, through the action of axionic
domain walls \cite{Huang:1985tt}, leaving
a final state with small-amplitude axionic fluctuations which evolve
adiabatically thereafter.
Under our assumptions, a mass prediction should be possible, because
the model has one principal free parameter,%
\footnote{\label{foot1}%
  There are two other relevant parameters.  There is the number $N_a$
  of minima around the U(1) circle,
  $\cos \argphi \to \cos\:N_a \argphi$.
  But if $N_a \neq 1$ then the model predicts stable domain walls
  which are a cosmological disaster
  \cite{Hiramatsu:2010yn,Hiramatsu:2012sc}.
  Also there is the mass of the radial excitation in the complex
  scalar field, $m$.  This must be heavy, and we find below that the
  results are quite weakly dependent on its exact value.}
the axion ``decay constant'' (the vacuum expectation value breaking
the U(1) symmetry) $f_a$ (defined below in \Eq{Lflat}).  This
parameter determines the axion mass, see \Eq{def_ma}.  And given our
other assumptions, it also sets the dark matter density.  By computing
the relation between $f_a$ and the dark matter density, we should then
be able to predict $f_a$ and therefore $m_a$.  Such a prediction is
valuable because it informs experiment, and because if the axion is
then discovered at this mass, it will clarify its role as the dark
matter.

In the next section, we lay out our methodology for relating $f_a$ to
the axionic dark matter density.  Section \ref{results} presents our
numerical results.  We end with a discussion.  A few technical issues
and numerical tests are postponed to an appendix.
But for the impatient reader, we present our main results here.
While there have been numerous previous studies of this problem
\cite{Harari:1987ht,Hagmann:1998me,Battye:1993jv,Battye:1994au,%
  Yamaguchi:1998gx,Yamaguchi:1999yp,Hiramatsu:2010yu,%
  Hiramatsu:2012gg,axion1,axion2},
ours is the first which includes the physics of the large tension
associated with axionic strings.  This large tension leads to a much
higher density of strings which are more robust and survive longer
than in previous simulations.  However, this makes surprisingly little
difference in the final axion number produced.
Let us set as a baseline for axion production, the angle-averaged
misalignment value of the axion density.
This is the axion density value we would find if the axion field
starts out uniform in space with value $\tA$, averaged over
$\tA \in [-\pi,\pi]$ (without the approximation, sometimes made, of
replacing $1-\cos\tA \to \tA^2/2$ in the potential)%
\footnote{\label{foot2}%
  Specifically, we write the conformal-time axion mass squared as
  $m_a^2=t^{n+2}/t_*^{n+4}$ ($t$, $n$, and $t_*$
  defined in the next section) and evolve $\theta(t)$ according to
  $d^2\theta/dt^2+(2/t)d\theta/dt = - m_a^2 \sin\theta$ to
  a time $t > 4t_*$.  The axion number density at time $t$ is
  $\nax = (m \theta^2 + \dot\theta^2/m)/2$.  We average over starting
  values $\theta \in [-\pi,\pi]$ and use the resulting $\nax$ average
  to normalize the result of a string simulation at the same $t/t_*$
  value.}.
At a given $f_a$ value, we find the axion number density produced in
the inhomogeneous case is actually \textsl{smaller} than the
misalignment value, by a factor of about 0.78.  Since $\rhodm$
increases with increasing $f_a$, this inefficiency must be compensated
by a larger value of $f_a$, and hence a smaller value of $m_a$,
than has generally been assumed; we find
$m_a \simeq 26.2 \pm 3.4 \:\mu\mathrm{eV}$.  We postpone discussion of
this result and its errors to the conclusions.

\section{Methodology}
\label{sec:method}

Our approach will be as follows.  The Lagrangian for the axion field is%
\footnote{%
  \label{foot3}We use $[{-}{+}{+}{+}]$ metric convention, and standard
  complex-field normalization $\varphi = (\varphi_r+i\varphi_i)/\sqrt{2}$.}
\begin{equation}
\label{Lflat}
  -\LL =  g^{\mu\nu} \partial_\mu \varphi^* \partial_\nu \varphi
  + \frac{m^2}{8f_a^2} \Big( 2 \varphi^* \varphi - f_a^2 \Big)^2
  + \chi(T) \left( 1 - \cos \tA \right) \,.
\end{equation}
The middle term is the symmetry breaking Lagrangian, and the last term
is the ``tilt'' in the potential due to QCD effects.  This tilt gives
rise to an axion mass of
\begin{equation}
  \label{def_ma}
  m_a^2(T) = \frac{\chi(T)}{f_a^2} \,, \qquad
  m_a^2(T{=}0) \, = \, \frac{\chi(T{=}0)}{f_a^2} \,.
\end{equation}
In a radiation
dominated FRW universe, in comoving coordinates and conformal time,
the metric is $g_{\mu\nu} = t^2 \eta_{\mu\nu}$ and the temperature is
$T \propto t^{-1}$, so
\begin{equation}
\label{Lconformal}
  -\sqrt{g} \, \LL_{\mathrm{conf}} = t^2 \left(
  -\dot\varphi^* \dot\varphi + \nabla \varphi^* \cdot \nabla \varphi
  + t^2 \frac{m^2}{8f_a^2} \Big( 2\varphi^*\varphi - f_a^2 \Big)^2
  + t^2 \chi(t) \left( 1 - \cos \tA \right) \! \right) .
\end{equation}
Here $\chi(T(t))$ is the topological susceptibility.  Model calculations
\cite{Wantz:2009mi} and a recent lattice calculation
\cite{Borsanyi:2016ksw} indicate that $\chi(T)$ is approximately power law
between 1.5 GeV and 400 MeV, which, we will see, is wider than the
relevant temperature range we need.  Therefore we will treat $\chi(T)$
as a power law, $\chi(T) \propto T^{-n}$, so
$t^2 \chi(t) = f_a^2 t^{n+2}/t_*^{n+4}$, with $t_*$ the natural scale where
the susceptibility begins to influence the dynamics;
$t_* m_{a,\mathrm{conf}}(t_*) = 1$
where $m_{a,\mathrm{conf}} = \sqrt{t^2\chi(t)} / f_a$ is the
conformal-time axion mass.  In terms of physical time, $t_*$ is the
moment when $m_a H=1$.  In the following we will suppress the
subscript and write $m_{a,\mathrm{conf}}=m_a$, except in the
discussion.  That is, masses and times will always be expressed in
conformal units.

To initialize the network, we choose an independent random phase at
every lattice site.  We then evolve the fields for an initial time
under strong damping
($\ddot\varphi+2\dot\varphi/t \to \ddot \varphi + \kmax \dot\varphi/t$
for times $t < \tstart$) to prepare a string network relatively close to
the scaling network density.  The length and strength of damping is
chosen such that the string network will roughly match on to the scaling
network density; we will also study the dependence on the initial
conditions below.

The model has two sorts of metastable defects, strings and domain
walls.  A string is identified as a linear structure where $\tA$
changes by $2\pi$ in circling the string.  A domain wall is a
surface on which $\tA = \pi$; each string has one such domain wall
ending on it.  The domain walls only become distinct structures once
$m_a t \gg 1$; the surface tension of an isolated domain wall is
$\sigma = 8m_a f_a^2$, which grows rapidly with time.  Therefore the
domain walls straighten out and pull the strings together,
annihilating both networks and leaving small fluctuations in the $\tA$
field.  We evolve all fields until this dynamics is complete and there
are no strings left.  Then we count the axion abundance by extracting
$\tA$ and $d\tA/dt$ from the simulation and applying the method of
\cite{axion1} to determine the axion content.  This determines the
total density of axions from all sources -- we make no attempt to
distinguish which axions arise from strings, from walls, or from
misalignment, as we do not believe such a distinction can be made
unambiguously.  We express the
axion number produced as a ratio to the angle-averaged misalignment
value and we determine $f_a$ such that the dark matter abundance is
correct.\footnote{\label{foot4}%
  We implement the misalignment case in the same code by turning off
  the scalar gradient terms.  We also implemented misalignment in a
  simple dedicated code as a cross-check.}

The scale $m$ for ``radial'' excitations in the $\varphi$ field
is may be as large as $m \sim f_a \sim 10^{11}\mathrm{GeV}$ and must be at
least $10^3 \:\mathrm{GeV}$ (see Subsection \ref{seckappa}),
while the relevant length scale is $H$ at the QCD epoch (around 1 GeV
temperature), which is of order $10^{-18} \mathrm{GeV}$.  We handle
this huge scale hierarchy by observing that the only important physics
it gives rise to is very thin, high-tension axionic cosmic strings.
Specifically, the string tension should be
$T_{\mathrm{str}} \simeq \kappa \pi f_a^2$ with
$\kappa \equiv \ln(m/H) \in [50,70]$.
We address this physics via the technique we recently introduced
\cite{axion3}.  Specifically, we add abelian-Higgs degrees of freedom
which are massive away from string cores but which induce a large
string tension.  In the Appendix \ref{sec:algorithm} we review the
procedure and explain how we implement $\chi(T) (1{-}\cos\tA)$ into
this method.  The outcome is that the string tension is maintained by
some extra, massive degrees of freedom, but the mass scale $m$ for
these degrees of freedom must be resolved by the lattice,
$ma \lesssim 1$ with $a$ the lattice spacing.  The correct physical limit
involves this scale becoming heavy compared to the physics of IR
fluctuations in the axion field.  Since the correct physical picture
arises when $m$ is large, we will hold $ma$ fixed, that is, we keep
$m$ fixed in lattice units, throughout a simulation.  In our
implementation, the extra degrees of freedom introduce one new
parameter $q_1$, which determines the value added to $\kappa$; most of
our results are labeled $(q_1,q_2)=(4,3)$, which means that $\kappa$
has been increased by 50 through the added degrees of freedom
\cite{axion3}.

\section{Numerical results}
\label{results}

In Appendix \ref{sec:test} we show that our results are in the large
volume limit if we keep $L/t_* \geq 5$, and we show that the axion
number does not evolve after the string network is gone and can be
measured as soon as no string is left, without concern that it will
evolve further.  Therefore box size and axion number measurement do
not contribute to our error budget.  Here we will instead
focus on those effects which still do.

\subsection{Lattice spacing}
\label{secspacing}

\begin{figure}
  \includegraphics[width=0.47\textwidth,bb=27 161 546 590]{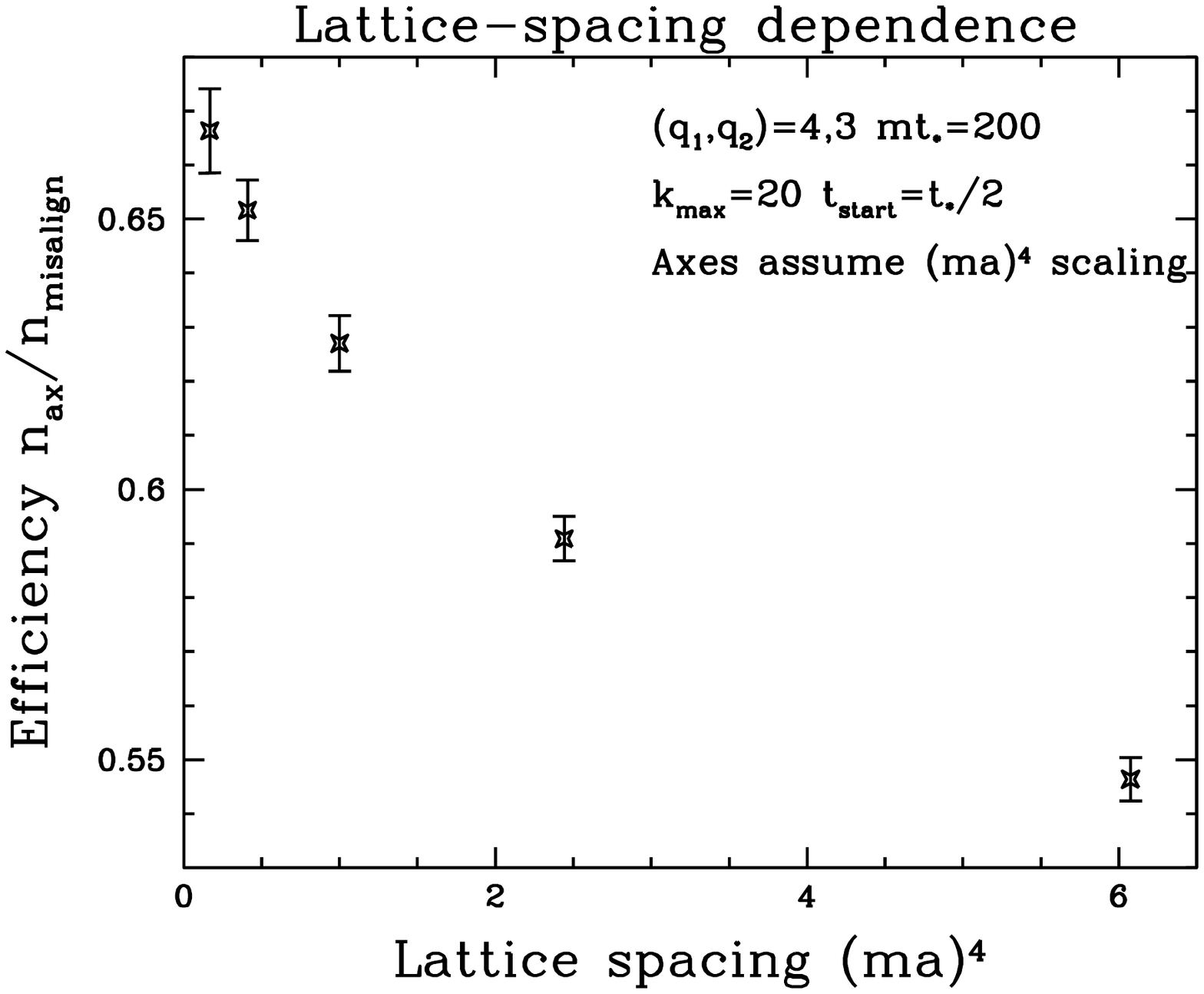}
  \hfill
  \includegraphics[width=0.47\textwidth,bb=27 161 546 590]{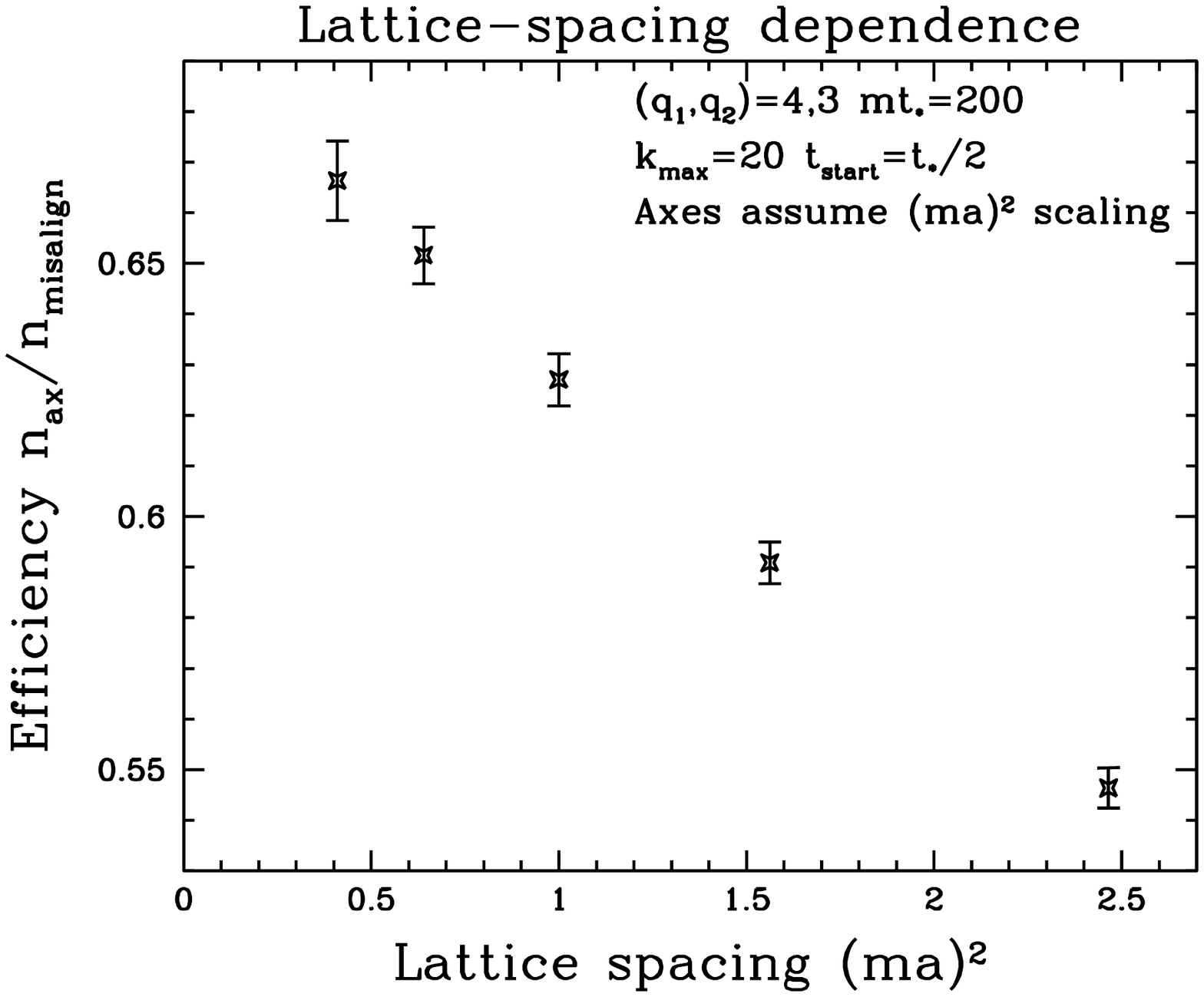}
  \caption{\label{fig:ma}
  Dependence of the axion production rate on the lattice spacing $a$,
  expressed as a function of $ma$ with all else held fixed in units of
  $m$.}
\end{figure}

For axionic strings to evolve correctly, the string core must be
resolved by our lattice spacing.  We need to check that our lattice
is fine enough, in the sense of $ma$ the product of the heavy scale
and the lattice spacing is sufficiently small.  That said, numerical
cost scales as $(ma)^{-4}$ and required RAM scales as $(ma)^{-3}$, so
we want the largest value which we can get away with.  To test the
$(ma)$ dependence, we fix all other parameters in terms of $m$, and
we consider axion production at various $ma$
values in Figure \ref{fig:ma}.  Because we have used an improved
action, the result should naively converge in the small $ma$ limit with
corrections vanishing as $(ma)^4$, motivating the axis choice in the
left-hand plot.  However, the right-hand plot shows that the data fit
better assuming $(ma)^2$ dependence.  Indeed the $\chi^2$ for an
$(ma)^2$ fit is about 1, while for an $(ma)^4$ fit, $\chi^2=15$.  Therefore
we will assume that the errors are quadratic in spacing, despite our
improved action.  The fit indicates an upwards correction between
$(ma)=1$ and the continuum limit $(ma)=0$ of
$10\pm 1\%$.  In the rest of this study we will use $ma=1$ and correct
the final results upwards by $10\%$.

It remains to explain why the axion production scales with $(ma)^2$
despite our improved action, which should give $(ma)^4$ convergence
if fields are smooth.  We believe this occurs because a small fraction
of string has a velocity close to 1, and therefore a large Lorentz
contraction factor.  If the (energy-weighted) fraction of string with
velocity-squared $v^2 > v^2_0$ only vanishes linearly in the
$v^2_0\to 1$ limit, then the fraction of string with
$\gamma^{-2} < \epsilon$ would then scale linearly in $\epsilon$.
Such scaling is consistent with our measured string velocity
distribution.  It also makes sense from the string equations of motion.
In flat space, with $\chi=0$, and in the Nambu-Goto limit, labeling
the string location as $x_i(\sigma,t)$ with $\sigma$ an affine
parameter along the string, one can make a gauge choice such that
$\dot{x}_i x_i'=0$ and such that
$\sqrt{x_i' x_i'/(1 - \dot{x}_j \dot{x}_j)}=1$.
The equation of motion is then
\begin{equation}
  \ddot{x}_i = x_i''
\end{equation}
which is solved by \cite{Vilenkin:1981sd,Turok:1984db,Bennett:1989yp}
\begin{align}
  x_i' & = \frac{\alpha_i(\sigma{+}t)+\beta_i(\sigma{-}t)}{2} \,,
  \\
  \dot{x}_i & =  \frac{\alpha_i(\sigma{+}t)-\beta_i(\sigma{-}t)}{2} \,,
  \\
  \alpha_i(r) \alpha_i(r) & = 1 = \beta_i(r) \beta_i(r) \quad
  \forall r \,.
\end{align}
Here $\alpha,\beta$ are backwards and forwards propagating waves which
take values on the unit sphere.  Even in curved space we may satisfy the
gauge choice instantaneously.  The relevant question for the
distribution of string velocities is the distribution of angles
between $\alpha$ and $\beta$, since
$v^2=\dot{x}^2 = (1-\alpha\cdot \beta)/2$.  The \textsl{measure} of
$\alpha\cdot \beta$ values is uniform in $[-1,1]$ because
$\alpha,\beta$ take values on the unit sphere.  While we do not expect
the distribution of $\alpha\cdot \beta$ values to be uniform in
$[-1,1]$, neither do we have a reason why it should avoid
$\alpha\cdot \beta=-1$, so the probability distribution should not
vanish at $\alpha\cdot \beta=-1$, and therefore the fraction of string
with $\gamma^{-2} < \epsilon$ should indeed vanish linearly in small
$\epsilon$.  Consider $\epsilon = (ma)^2$, corresponding to a
gamma-factor of $\gamma > 1/ma$ and therefore a Lorentz contracted string
thickness%
\footnote{\label{foot5}%
  In this parametric argument we are neglecting order-1 factors which
  make the string somewhat thicker than $1/m$ and mean that, for
  $ma=1$, \textsl{most} string is actually properly treated.}
of $\gamma^{-1}(1/m) < a$.  Our hypothesis for string velocities
then states that an $\OO((ma)^2)$ fraction of string should be Lorentz
contracted to a thickness of less than 1 lattice spacing.
Such string is mistreated regardless of how improved our update
algorithm is.  Therefore, even if typical string is treated correctly
with $\OO((ma)^4)$ errors, the fraction which is mistreated is of
order $(ma)^2$.  This allows $(ma)^2$ scaling corrections, regardless
of the level of lattice action improvement.

\subsection{Initial network density}
\label{secdens}

We want the axion production from a string network which is initially
in the scaling regime.  But this cannot be exactly achieved; initial
conditions will typically produce a network which is either denser or
less dense than scaling.  The network evolves towards scaling, and if
$mt_*$ is large enough then initial conditions should have little
effect.  But it would still be good to check how sensitive the final
axion number is to the starting conditions.

\begin{figure}
  \hfill
  \includegraphics[width=0.65\textwidth,bb=26 157 557 588]{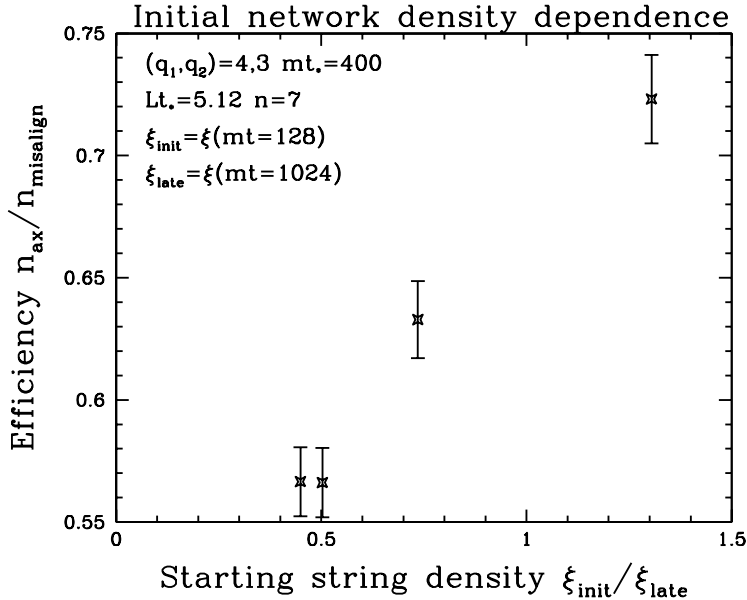}
  \hfill $\phantom{.}$
  \caption{\label{fig:init}
    Dependence of the final axion density on the initial string
    network density.  We measure the starting network density as the
    ratio between the scaled density $\xi$ at time $mt=128$
    ($t/t_* = 0.32$) and the scaled density at time $mt=1024$, shortly
    before the walls start to influence the network evolution
    ($t/t_* = 2.56$).}
\end{figure}

We address this in Figure \ref{fig:init}.  We introduce the scaled
network density
\begin{equation}
  \label{def_xi}
  \xi = \frac{t^2 \int_{\mathrm{all\;string}} \gamma dl}
      {4V_{\mathrm{space}}} \,,
\end{equation}
with $\gamma$ the local gamma-factor of the string so that the
integral represents the total invariant length of string (length
scaled by a $\gamma$ factor to account for the energy content), and
with $V_{\mathrm{space}}$ the volume of the simulation.  This combination
should approach a fixed ``scaling'' value as $t$ increases (for
$\chi(T)=0$, that is, in the absence of potential tilt).  We measure
$\xi$ once early in an evolution and again later in the evolution, just
before the walls start to influence the string network evolution.  We
perform several evolutions with initial conditions with more or less
damping, leading to denser or rarer initial networks.  The ratio of
the starting to final $\xi$, $\xi_{\mathrm{init}}/\xi_{\mathrm{late}}$,
then indicates whether the network started too thin or too dense, and
therefore from which side it is approaching the scaling solution.  We
find a roughly linear correlation between this starting density and
axion production, with more axions arising from denser starting
networks.  However the dependence is quite weak.  Based on the figure,
we will try to use initial conditions with this $\xi$ ratio close to
1, and we will assign a 5\% systematic error based on incomplete
network scaling.

\subsection{Thin-core limit}
\label{secthin}

Next we must consider the effects of finite $mt_*$, meaning that the
strings are of finite thickness.  This is clearly an artifact because
in the physical case there is a hierarchy of many orders of magnitude
between string thickness and axion mass.  We have incorporated the
logarithmic sensitivity to this hierarchy by implementing auxiliary
fields to give rise to the resulting high string tension.  But there
can still be effects suppressed by powers of $1/(mt_*)$, probably
starting at first order.  In particular,
strings may lose energy via the radiation of unphysical massive
modes.  We only expect such radiation from short length-scale
structures on the strings, which should generally get smoothed out by
axion emission so long as $mt_* \gg \kappa$ \textsl{and}
$m_a/m \ll 1$. However, because $m_a$ grows as a large power of
conformal time, $m_a \propto t^{1+\frac n2}$, the latter condition may not
be maintained, given the persistence of high-tension strings.
And if the axion mass $m_a$ comes of order the heavy-mode mass $m$
then one might expect that axion production is lost to heavy-mode
production, and the simulation could result in an
\textsl{under}estimate of axion production.%
\footnote{\label{foot6}%
  In previous work \cite{axion1} we showed that, for a theory of a pure scalar
  field, the axionic domain walls spontaneously decompose as soon as
  $(m_a/m)^2 > 1/39$.  The added degrees of freedom in our string
  cores prevent this physics from occurring; the domain walls remain
  strongly metastable up to and past $m_a/m = 1$.}

\begin{figure}[htb]
  \hfill
  \includegraphics[width=0.65\textwidth,bb=48 162 568 612]{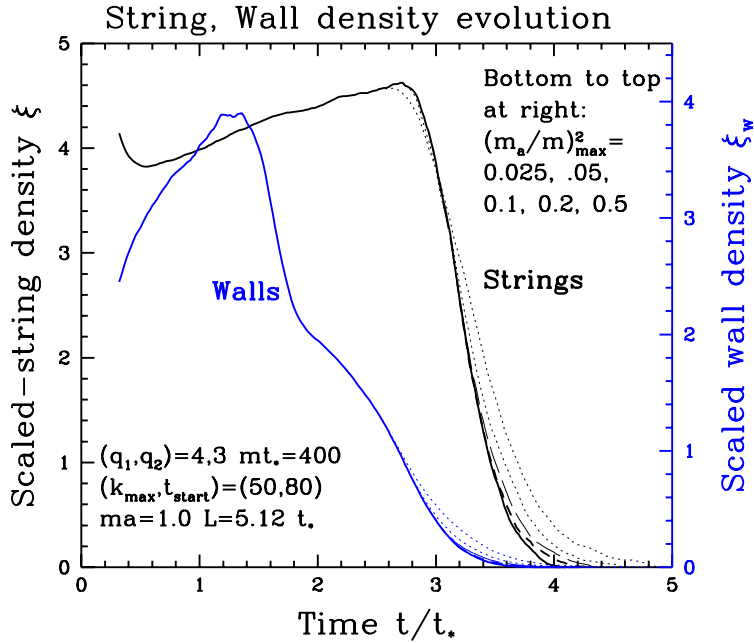}
  \hfill $\phantom{.}$
  \caption{\label{fig:network}
    Density of walls and of strings as a function of $t/t_*$ when the
    axion mass is artificially limited with 5 different limits.}
\end{figure}

We can ``fix'' this problem by artificially
capping the value of $m_a$, so that rather than growing with time at
all times, $t^2 \chi(t)$ grows up to some value and then becomes
constant.  But this replaces one unphysical behavior with another, and
it will introduce new artifacts.  The axionic wall tension is
proportional to $m_a$.  These walls cause the network to collapse, and
limiting their tension artificially extends the life of the network.
We show this effect in Figure \ref{fig:network}.
The figure shows how the total length of strings, rescaled as in
\Eq{def_xi}, and a similar rescaled wall area (without $\gamma$ factor
or the conventional factor of 4),
\begin{equation}
  \xi_{\mathrm{wall}} \equiv \frac{t \int_{\mathrm{all\:wall}} d^2\Sigma}
     {V_{\mathrm{space}}} \,,
\end{equation}
evolve with time under the influence of various choices for a maximal
$m_a/m$ value.  We see that the wall area starts to decline as the
wall surface tension turns on around $t=1.6 t_*$, and later around
$t=2.8t_*$ the surface tension becomes large enough to influence the
string network evolution, drawing together the strings and collapsing
the network by $t=4t_*$. However, artificially limiting the axion mass
slows down the collapse of the network; for the smallest value we
considered, the last bits of string survive almost to $t=5t_*$.

We also expect that the network with a maximal $m_a/m$ value will produce
more axions than without such a cutoff.%
\footnote{\label{foot7}%
  Something special happens if we choose a maximum value for $m_a$
  which is very close to $m/2$.  In this case, there is a resonant
  nonlinear mode-coupling process which converts mass $m/2$ axions
  into mass-$m$ excitations, leading to a reduction in the axion
  production for values of $(m_a/m)_{\mathrm{max}}$ very close to 0.5.
  This effect is clearly an artifact, so we avoid this special value.}
The reason is that, as
$m_a$ increases, the energy stored in the string network becomes less
and less useful for producing axions.  While increasing $m_a$
increases the energy in axion fluctuations and in domain walls in
proportion to $m_a$, it does not change the energy in strings.
Therefore, as $m_a$ increases, the capacity for strings to make axions
is diluted; since making an axion costs energy $m_a$, an energy
$E$ can only produce $E/m_a$ axions.  Limiting
$m_a$ turns off this dilution, allowing the string energy to produce
more axions, and could therefore result in \textsl{over}production of
axions.

\begin{figure}[htb]
  \hfill
  \includegraphics[width=0.65\textwidth,bb=29 163 548 596]{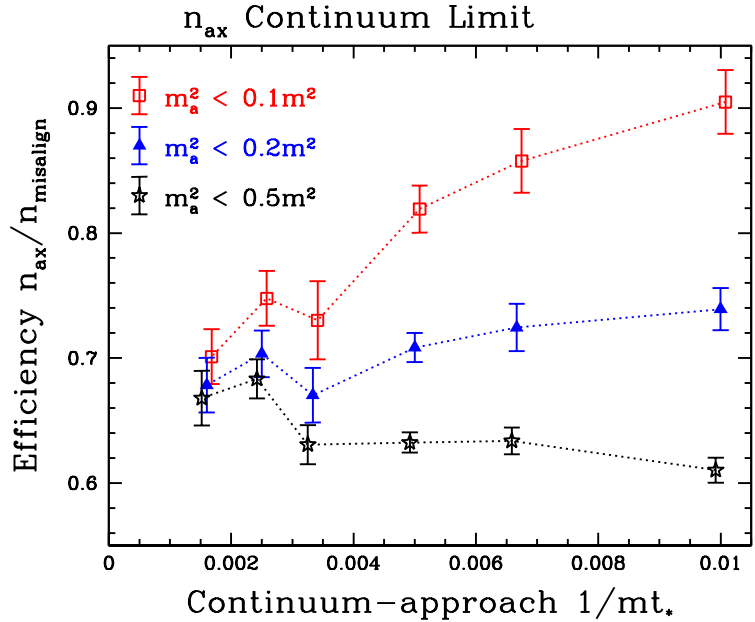}
  \hfill $\phantom{.}$
  \caption{ \label{fig:Lmax}
    Axion production as a function of the continuum limit
    $1/mt_* \to 0$ at three values of the artificial limiting value
    for the axion mass $(m_a/m)_{\mathrm{max}}$.}
\end{figure}

The issue should disappear if we can reach a large enough value of
$mt_*$.  But it is useful to consider different cutoff values for
$m_a/m$ and take the large $mt_*$ limit for each.  If the continuum
limits are the same, then it lends credence to the belief that we have
achieved the continuum limit.  According to Figure \ref{fig:Lmax},
The difference between different cutoff choices falls below 10\%
starting around $mt_*=300$ (third-from-leftmost points).  On the two
still-finer lattices, the choices $\mrsq=0.5$ and $\mrsq=0.2$
agree to within 3\%.  So these values can be close to the continuum
limit.  Note that the last point in
the figure, with $mt_*=625$, was achieved by loosening $ma$ from
$ma=1.0$ to $ma=1.25$ and using the result of Figure \ref{fig:ma} to
extrapolate it to the same value as the other points; it is also at a
slightly smaller physical volume, $Lt_*=4.1$ rather than 5.1.

The error bars shown in the figure are statistical only.  However the
statistical errors for points with the same $mt_*$ value but different
$\mrsq$ values are strongly correlated, since they are calculated
from simulations which are identical up to the point when $m_a$
reaches the smaller upper-bound value.
Therefore the determination of the difference
between different $\mrsq$ choices has smaller errors.  In
particular this difference is \textsl{not} linear in $1/mt_*$, but
drops to a small value at a sufficient $m_{a,\mathrm{max}}t_*$ value.
That complicates the continuum limit.  Here we will perform a linear
extrapolation of the three smallest $1/mt_*$ data points, each
for $\mrsq=0.2$ and $0.5$.  We find
$0.696(46)$ and $0.729(41)$ respectively.  The fact that these answers
are not the same indicates that our lattices are not yet abundantly
fine.  We assign a 10\% error bar for the continuum-extrapolated
value, to include these systematic issues, adopting
$\nax/\nmiss = 0.71(7)$.
This is for $n=7$,
$\kappa=50$ from extra degrees of freedom, and before performing the
small $ma$ extrapolation.

\subsection{String tension and temperature-dependent susceptibility}
\label{seckappa}

Having discussed numerical artifacts, we now turn to actual physical
parameters which are relevant but not completely known:  the string
tension $\kappa$ and the strength of the temperature dependence $n$ in
$\chi(T) \propto T^{-n}$.

\begin{figure}[htb]
  \hfill
  \includegraphics[width=0.65\textwidth,bb=21 160 547 592]{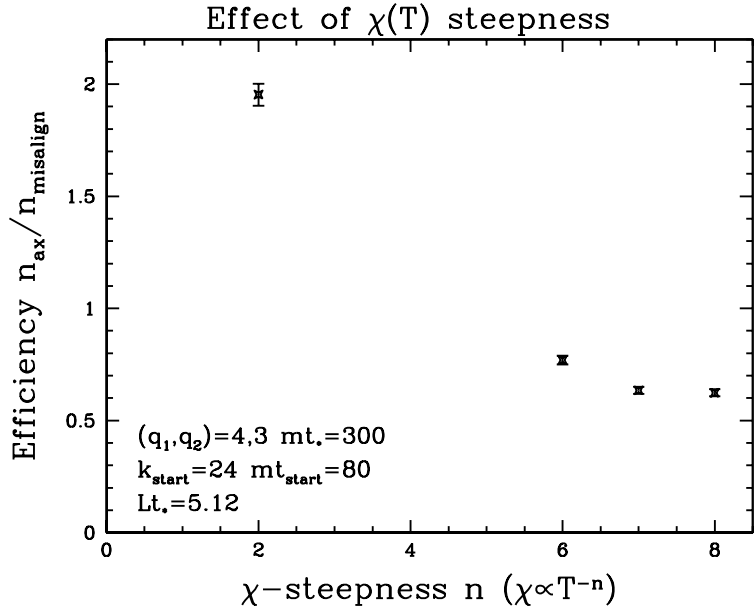}
  \hfill $\phantom{.}$
  \caption{\label{fig:n}
    Dependence of the relative axion production efficiency on the
    parameter $n$, controlling how quickly $m_a$ rises with $t$.  For
    small $n$ (gradually rising $m_a$) the network produces more
    axions than in misalignment; for large $n$ it produces fewer.
  }
\end{figure}

The slope $n$ is calculable in lattice QCD.  Recently Borsanyi
\textsl{et al} have presented \cite{Borsanyi:2016ksw} results up to and
beyond the relevant temperature range.  Using their results at 600 and
1200 MeV, we estimate $n=7.6 \pm 0.5$.  Most groups find results which
agree with Borsanyi \textsl{et al} at lower temperatures
\cite{Berkowitz:2015aua,Borsanyi:2015cka,Petreczky:2016vrs,
  Taniguchi:2016tjc,Burger:2017xkz,Frison:2016vuc},
although no group has reproduced these higher temperatures and even
below 600 MeV there are some results which appear discrepant
\cite{Bonati:2015vqz,Bonati:2016imp}.
Therefore we will explore other $n$ values but consider values near
$n=7.5$ to be likely correct.  We also feel that we gain some physical
insight by considering different $n$ values, especially much smaller
values.  We do this in Figure \ref{fig:n}.
The figure shows that small $n$ values lead to more axions than in the
misalignment mechanism, while large $n$ values lead to less.  But
between $n=7$ and $n=8$ the dependence is not very strong.  Therefore
our choice to use $n=7$ elsewhere, which we made mostly for
simplicity, does not appear to be very critical.

We interpret the results of Figure \ref{fig:n} as follows.  The larger
the $n$-value, the more rapidly the axion mass $m_a$ turns on, and
therefore the heavier the axion is when the string network breaks up
and loses its energy.  That means that for small $n$, the network can
still produce relatively many of the relatively-light axions, but for
large $n$ the axions quickly become heavy and the string energy cannot
produce a large number of them.  This is consistent with what we saw
when $\mrsq$ was small.  Indeed, the results at
$n=2$ had $(m_a/m)^2=0.07$ at the time the string network had
completely disappeared, so walls broke up and axion production
occurred when axions were still relatively light.

Finally we consider the $\kappa$ value.  Above we define $\kappa$ as
$\kappa = \ln(m/H)$.  For us $H = 1/t$ the inverse system age.
Therefore the contribution from axionic modes to $\kappa$ is
$\ln(mt)$, which we approximate to its value at $t=3t_*$ since this is
when the string network is breaking up.  In addition there is a
contribution from the extra massive degrees of freedom we have added,
so our simulations have
\begin{equation}
  \label{kappa_eff}
  \kappa = \ln(3mt_*) + 2(q_1^2 + q_2^2) \,,
\end{equation}
where the charges $(q_1,q_2)$ are explained in Appendix
\ref{sec:algorithm}.

We do not know what the physical value of $\kappa$ should be, because
we don't know the model-dependent microscopic origin of the axion
field.  In the single complex-field case
\cite{Kim:1979if,Shifman:1979if} we don't know the radial mass $m$; if
the axion is a composite or arises from more complicated physics
\cite{Kim:1984pt}, we do not know the compositeness scale and whether
there is an extra contribution to the string tension from the
microscopic physics giving rise to the axion field.  We can reasonably
guess that $m < f_a \simeq 2\times 10^{11}\:\mathrm{GeV}$.  Also the
requirement that the radial excitations decay by the time the Universe
reaches a temperature of 1 GeV, along with an estimate for their decay
rate \cite{Fox:2004kb},
$\Gamma_m \sim \frac{\alpha_s^2}{64\pi^3} \frac{m^3}{f_a^2}$, sets very
roughly $m > 10^3\:\mathrm{GeV}$.  These limits correspond to
approximately $\kappa \in [48,67]$.  For $mt_*=300$ and $q_1=4$ the
$\kappa$ value in the simulation is $50+\ln(900)=57$, which is in this
range.  By considering other values of $(q_1,q_2)$, we achieve
$\kappa$ values larger and smaller than the physically interesting
range.

\begin{figure}[htb]
  \includegraphics[width=0.49\textwidth,bb=46 160 550 592]{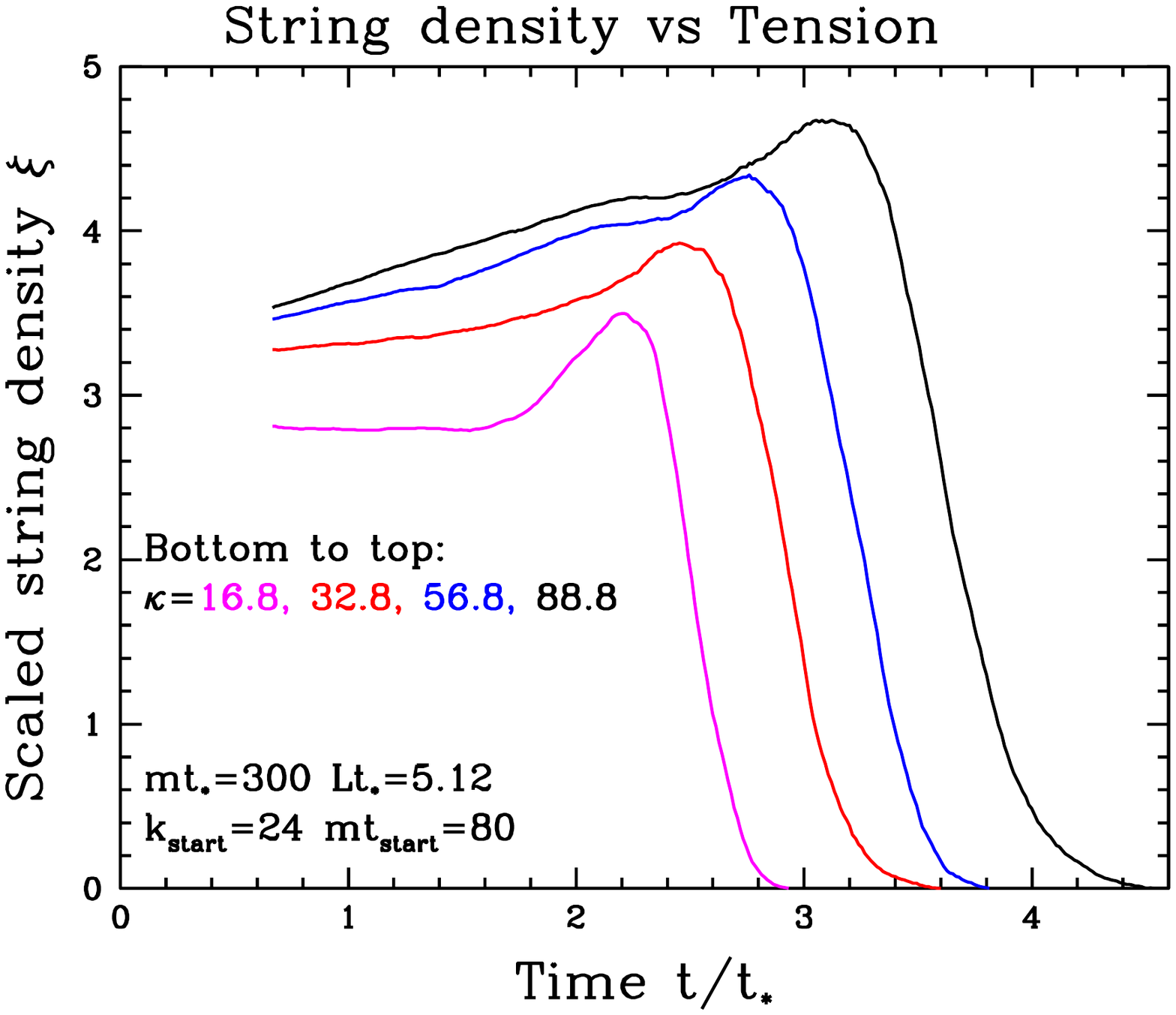}
  \hfill
  \includegraphics[width=0.49\textwidth,bb=21 143 558 591]{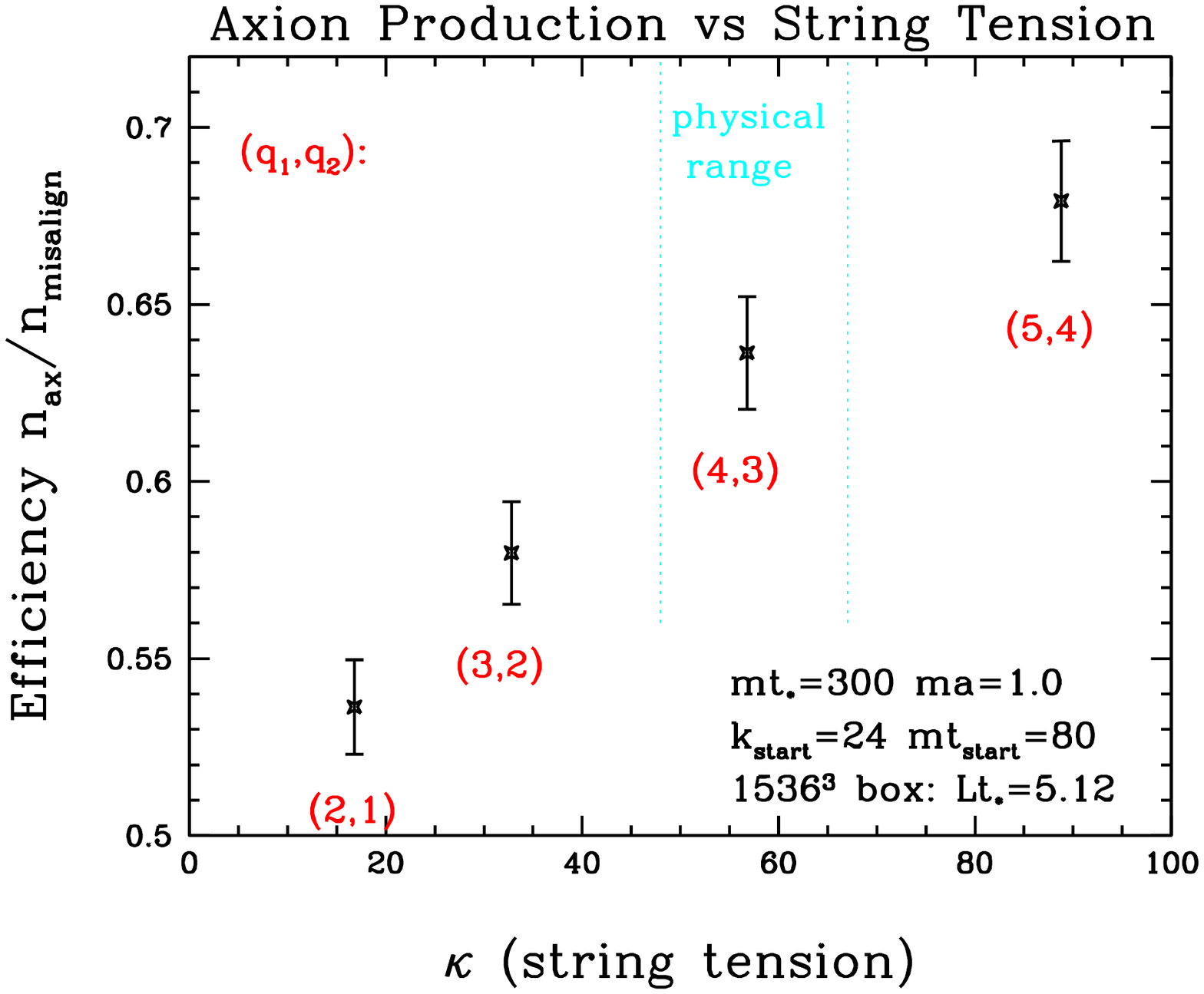}
  \caption{\label{fig:kappa}
    Left:  string density as a function of time for different $\kappa$
    values.  The higher the string tension, the longer the strings
    persist.  Right:  axion production efficiency as a function of
    $\kappa$.
  }
\end{figure}

We show results for the axion production as a function of $\kappa$ in
Figure \ref{fig:kappa}.  The figure shows that higher tension strings
give a significantly denser string network, with strings which break
up later, but nevertheless produces only mildly more axions (note the
false zero for the y-axis in the right plot).
Therefore our ignorance of the physical value of $m,\kappa$ is not very
significant in bracketing the physical value of the axion.  A simple
linear fit to the left frame in Figure \ref{fig:kappa}, and the range
we quoted above for $\kappa$, gives a systematic error of
$\pm 3\%$ due to the unknown value of $\kappa$.

Note that the chosen initial conditions for both our $n$ dependence
and our $\kappa$ dependence studies produced somewhat underdense
networks.  One can see in the left frame of Figure \ref{fig:kappa}
that the underdensity is worse for the highest tensions, so the true
$\kappa$ dependence is somewhat \textsl{under}estimated.  Also note
that the larger $\kappa$ values are farther from the large $mt_*$
limit, leading to a slight \textsl{over}estimate in the produced
axions due to $\mrsq$ effects.

\section{Discussion}
\label{discussion}

If we take the temperature dependence of the topological
susceptibility to scale as $\chi(T) \propto T^{-n}$ with $n=7.6$ in
the relevant temperature range \cite{Borsanyi:2016ksw}, and assume
that axionic string cores arise at a mass scale
$m \sim 10^7\:\mathrm{GeV}$ so the extra string tension is
$\kappa = 58 \pm 10$, then our results indicate an axion production
efficiency which is
$0.78(12)$
times as efficient as in the angle-averaged misalignment mechanism.
The indicated error is dominated by the extrapolation to the large
$mt_*$ limit, with the uncertainty due to $\kappa$
added linearly (not in quadrature).

Now we use this result to calculate the axion mass.  There has been
relatively little entropy production since the Universe was 1
GeV in temperature, so the ratio of axion number-density to entropy
density is approximately the same at the end of the axion-production
epoch as it is now.  We can express our results by
saying that the axion number density, determined at
$t=4t_*$ or $T=T_*/4$ and then back-extrapolated to the temperature
$T_*$ where $m(T_*) H(T_*)=1$, was
\begin{equation}
\mbox{back-extrapolated} \quad
  \nax(T=T_*) \simeq K H(T_*) f_a^2 \,,
\end{equation}
where a numerical evaluation finds that the angle-averaged
misalignment value for $K$ is $K=16.61$, and our result is
$K=13.0 \pm 2.0$.
By dividing this by the entropy density at that temperature,
$s=2g_* \pi^2 T_*^3/45$, we get the modern axion-number to entropy
ratio, which can be multiplied by the vacuum axion mass
$m_a = \sqrt{\chi(T=0)}/f_a$ to give the modern dark matter density to
entropy density ratio.
We combine this with the Planck result \cite{Ade:2015xua},
\begin{align}
\label{Planck}
\frac{n_b}{s} &\simeq 8.59\times 10^{-11} \,, \nonumber \\
\frac{\rhodm}{s} & = \frac{\Omega_{\mathrm{dm}} h^2}{\Omega_b h^2}
\frac{m_p n_b}{s} \simeq \frac{0.1194}{0.0221}(938\:\mathrm{MeV})
(8.59\times 10^{-11})
\simeq 0.39 \:\mathrm{eV},
\end{align}
thermal QCD results for the entropy density $s$ and energy density
$\varepsilon$ of the thermal plasma from Borsanyi \textsl{et al}
\cite{Borsanyi:2016ksw},
\begin{align}
  \varepsilon(T) & = \frac{\pi^2 T^4 g_*}{30} \,, \quad
  s(T) = \frac{2\pi^2 T^3 g_*}{45} \,, \qquad
  g_*(1\:\mathrm{GeV}) \simeq 73,
  \\
  \chi(T) & \simeq \left( \frac{1\:\mathrm{GeV}}{T} \right)^{7.6}
  (1.02(35)\times 10^{-11} \: \mathrm{GeV}^4) \,,
\end{align}
Hubble's law $H^2 = 8\pi \, \varepsilon/(3 m_{\mathrm{pl}}^2)$ with
$m_{\mathrm{pl}} \simeq 1.22\times 10^{19}\:\mathrm{GeV}$,
the thermal value for the axion mass $m_a^2(T) = \chi(T)/f_a^2$,
and vacuum value $\chi(T=0)=(0.076\:\mathrm{GeV})^4$
\cite{diCortona:2015ldu}, to obtain
\begin{align}
  \label{fa-result}
  f_a & = (2.21 \pm 0.29)\times 10^{11} \: \mathrm{GeV},
  \\
  \label{ma-result}
  m_a & = 26.2 \pm 3.4 \: \mu\mathrm{eV} ,
  \\
  \label{T*-result}
  T_* & = 1.54 \pm 0.05 \: \mathrm{GeV} \,.
\end{align}
Taking the errors quoted in Ref.\cite{Borsanyi:2016ksw} at face value,
the dominant error in $f_a$ and hence in $m_a$ is from our
determination of $K$, while the error in $T_*$ arises equally from the
errors in $K$ and in $\chi(T)$.  \Eq{fa-result}, \Eq{ma-result}, and
\Eq{T*-result} constitute the main results of our study.

Our most striking result is that the axion production from random
initial conditions, with the resulting dense and high-tension axionic
string network, is actually
\textsl{smaller} than the angle-averaged misalignment value.  The
deficit gets larger at large $n$, where the $\tA$ potential tilts more
abruptly; if it tilts more gradually then the axion production exceeds
the misalignment value.  Furthermore, although axion production is
larger from high-tension strings than from strings with a lower
tension, the dependence is quite weak; a factor of 10 increase in
string tension between our results and the results of \cite{axion1},
along with the resulting factor of 3 increase in the string network
density, has led to less than a $30\%$ increase in axion production.

This clearly requires some explanation.  The conventional wisdom has
been (see for instance \cite{Hiramatsu:2012gg})
that axions are produced by misalignment in the space between walls,
by walls, and
by strings.  Therefore the production is the sum of three terms, and
must be larger than the misalignment contribution.  We argue that this
picture involves assumptions and commits double counting.  It
does not make sense to consider misalignment axions to be independent
from walls.  Within the misalignment mechanism, half of all axions
emerge from the range of angles $|\tA(t=0)| \in [2.76,\pi]$.  But it
is precisely the regions with $\tA \sim \pi$ which become the domain
walls.  Much or most of the ``misalignment'' axion field energy
becomes the domain walls; it is double counting to speak of both
domain-wall axions and misalignment axions as independent
contributions.  Of course, since the axion field is initially
very inhomogeneous, it is also not obvious that there are \textsl{any}
spacetime regions where homogeneous misalignment is a useful
description. 

Consider also what happens to the energy in domain walls.  After the
potential tilts and the domain walls become relatively thin and
distinct, the wall surface tension induces forces on the strings.  The
walls lose their energy to accelerating the strings, which consumes
the wall area (see Figure \ref{fig:network}).  Also in this epoch, it
is not simple for walls or
strings to emit axions.  The axion frequency $m_a$ increases with
time, and any process involving time scales longer than $m_a^{-1}$ has
a frequency-mismatch problem to produce massive axions.  That is, long
wavelength fluctuations of walls or strings are incapable of producing
axions because they drive the axion field at frequencies below $m_a$.
We saw this very clearly in our previous study of 2+1D axion
production with massive strings \cite{axion2}.

What about the energy of the string network?  The high string tension
means that the network stores much more energy.  But after the time
scale $t_*$, the energy in
domain walls and in axionic fluctuations increases with the
axion mass as $E \propto m_a \propto t^{1+\frac n2}$, while the
string energy does not increase as $m_a$ increases.  Therefore the
string network's ability to produce axions dilutes with time.  The
network only annihilates when the walls are able to influence string
dynamics, which occurs when the wall energy is comparable to the
string energy.  That is, the strings only fragment when their energy
is comparable to the energy that was present in the wall network which
caused them to fragment.  And there is still the question of how
efficiently the resulting small loops turn their energy into axions.

To improve this analysis, we see a few directions which need to be
pursued.  First, we need simulations with more RAM, so that larger
boxes, and therefore larger $mt_*$ values, can be studied.  We need to
be more systematic in setting the initial network density and
understanding the approach to network scaling.  It should be
straightforward to reduce statistical and extrapolation errors to the
few percent level, with the dedication of more computer power.

Also, we
would like to investigate some of the late network evolution in more
microscopic detail.  The string network breaks up into loops which
then annihilate in a way which somehow does not produce many axions.
It should be possible to cut such loops out of a simulation and
resolve them with a much finer lattice, which can then properly
separate the $m_a$ and $m$ mass scales and follow the loop dynamics
down to short scales.  This could help explain why so few axions are
produced (or determine whether our limited lattice spacing is causing
a systematic neglect of some relevant but shorter-distance physics).

\section*{Acknowledgments}

We would like to thank Mark Hindmarsh for very useful conversations,
and Daniel Robaina, Thomas Jahn, and Max Eller.  We also thank the GSI
Helmholtzzentrum and the TU Darmstadt and its Institut f\"ur
Kernphysik for supporting this research.

\appendix

\section{Algorithmic details}
\label{sec:algorithm}

Here we explain in more detail how our numerics work.  Following
Ref.~\cite{axion3}, we embed the axion field as a global U(1) symmetry
of a theory with two scalars $\varphi_1$, $\varphi_2$ and one U(1)
gauge symmetry, so one linear combination of the U(1)$\times$U(1)
symmetry is gauged and one is global.  Both are spontaneously broken
by the scalar vacuum values:
\begin{align}
  \label{L-2field}
  - \LL(\varphi_1,\varphi_2,A_\mu) & =
  \frac{1}{4e^2} F_{\mu\nu} F^{\mu\nu}
  + \Big| (\partial_\mu -i q_1 A_\mu) \varphi_1 \Big|^2
  + \Big| (\partial_\mu -i q_2 A_\mu) \varphi_2 \Big|^2
  \nonumber \\ & \phantom{=} {} +
  \frac{m_1^2}{8 v_1^2} \Big( 2\varphi_1^* \varphi_1 - v_1^2 \Big)^2
  + \frac{m_2^2}{8 v_2^2} \Big( 2\varphi_2^* \varphi_2 - v_2^2 \Big)^2 .
\end{align}
Here $q_1$, $q_2$ are the field charges with $q_2=q_1-1$.
The axion is the angle
\begin{equation}
  \label{axdef}
  \tA = q_2 \argphi_1 - q_1 \argphi_2
\end{equation}
which is gauge-invariant.
This procedure exactly reproduces the global symmetry and the way
strings act as a source for the axion field; the dynamics are modified
only by a large induced string tension,
$\kappa \simeq 2(q_1^2{+}q_2^2)$, and heavy degrees of freedom which
should decouple from the dynamics in the continuum limit (in the sense
of $mt \gg 1$, or $mt_* \gg 1$ for our current purposes)\cite{axion3}.
We consider $v_1=v_2$ and
$m^2_1=m^2_2 = e^2(q_1^2 v_1^2 + q_2^2 v_2^2)=m_e^2$ so all heavy
fields have a common mass.  The axion decay constant is
$f_a^2 = v_1^2 v_2^2/(v_1^2 q_1^2 + v_2^2 q_2^2)$.  In our lattice
units we normalize our fields such that $v_1=1=v_2$.  The topological
susceptibility part of the potential (which breaks the global U(1)
symmetry, ``tilting'' the potential for the axion field), is
implemented as
\begin{align}
\label{chilatt}
  t^2 \chi(t) \Big( 1 - \cos \:\mathrm{Arg}\,\tA \Big)
  & \Rightarrow \frac{f_a^2 t^{n+2}}{t_*^{n+4}}
  F(2\varphi_1^*\varphi_1) F(2\varphi_2^* \varphi_2)
  \left( 1 - \cos \Big( q_2 \argphi_1 - q_1 \argphi_2 \Big) \right) 
  \\
\label{Flatt}
  F(r) & \equiv \left\{ \begin{array}{ll}
    \frac{25}{16} r \left( \frac{8}{5} - r \right)\,, \quad
    & r< \frac 45 \,, \\
    1\,, & r > \frac 45 \,. \end{array} \right.
\end{align}
The function $F(r)$ is inserted to soften the behavior of the
susceptibility term in string cores; without this term the introduced
potential becomes violently nondifferentiable wherever
$\varphi^* \varphi \sim 0$ for either field, which causes problems for
space-discretized equations of motion.  The modification only changes
the tilted potential inside string cores, where its effect is very
subdominant to the leading potential terms.  But without this
modification we do not get consistently stable evolution near string
cores.  Our results are insensitive to the specific form of $F(r)$,
provided $F(1)=1$, $F'(1)=0$, $F(0)=0$, and $F'(r)$ is continuous,
which motivated our choice.  A similar modification is common in
single-scalar simulations of axionic strings, where most authors
\cite{Hiramatsu:2010yn,Hiramatsu:2012gg}
have made the substitution
$(1-\cos\argphi) \to \sqrt{2}\:\mathrm{Re}\: \varphi$, a substitution
which is correct only for the angular dynamics and only where
$2\varphi^* \varphi = 1$.  This replacement is justified because it is
simpler, is nonsingular at 0, and is only of much influence outside
string cores, where it is nearly equivalent to the correct form.  We
have explicitly checked that in the single-scalar model, axion
production and string dynamics are nearly indistinguishable whether we
use $(1-\cos\argphi)$ or $\sqrt{2}\:\mathrm{Re}\:\varphi$ as the
``tilt'' in the potential.

Our numerical implementation uses a standard leapfrog algorithm and
the noncompact formulation of U(1).  The only novel feature is that we
use an $a^2$-improved action for both the scalar and gauge parts,
which requires a somewhat nontrivial treatment of electric fields in
which the link's canonical momentum is not the same as the link's time
derivative \cite{Moore:1996wn}.  More details and tests are in
\cite{axion3}.

\section{Other numerical tests}
\label{sec:test}

Here we detail some tests which have little bearing on the
extrapolation to a final result, and which we have therefore not put
in the main development.

\begin{figure}
  \hfill
  \includegraphics[width=0.65\textwidth,bb=28 164 557 610]{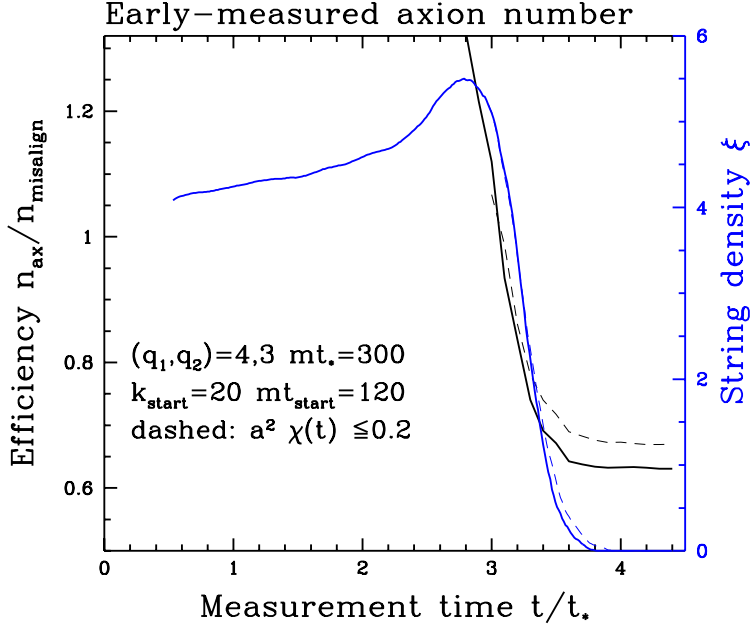}
  \hfill  $\phantom{.}$
  \caption{\label{fig:measearly}
    Blue: scaled string density $\xi$ as a function of time.  Black:
    instantaneously measured axion number at the same time.  Dashed
    lines are for an evolution with an upper cutoff on $\chi(T)$ at
    $\chi(T)<m^2/5$.  }
\end{figure}

In the main text we spend some effort considering when to stop the
growth of $\chi(T)$.  But we do not discuss when to measure the axion
number, arguing only that it is sufficient to measure after the string
network is gone and only small fluctuations remain.  Here we justify
this claim.  Figure \ref{fig:measearly} shows what happens when we
measure the axion number before the string network has finished
collapsing.  The figure shows the density of strings in blue, and the
density of axions, as measured at the indicated time, in black.  This
measurement is somewhat ambiguous because it involves identifying the
axion angle $\tA \in [-\pi,\pi]$ which is discontinuous across domain
walls.  Such a discontinuity leads to ``ringing'' in the Fourier
spectrum and formally gives a logarithmically UV divergent particle
number (cut off by lattice effects).
We ``fix'' this problem by truncating the largest $\tA$ values,
reflecting $\tA \in [\pi/2,\pi]$ to
$\tA \to \pi - \tA$, and similarly for $\tA \in [-\pi,-\pi/2]$.
Despite this ``cap'' on the maximum size of $|\tA| < \pi/2$, we
nevertheless find a very large axion density if we measure axions
before the network has decayed.  However, we see that after the
strings are gone, the axion number becomes completely independent of
further time evolution.  Our ``cap'' on large $\tA$ values has no
effect on this final plateau, because $|\tA| > \pi/2$ virtually never
happens and represents a tiny fraction of the axion number.
In light of this result, we
generally measure $\nax$ as soon as no strings remain, but when we
evolve for longer and remeasure later, we get an answer which agrees
at the $1\%$ level.
If we repeat this analysis for the misalignment scenario, we find that
instead of becoming virtually $t$-independent at $t=4t_*$, the axion
number becomes virtually $t$-independent already by $t=2t_*$.  This
difference reflects the absence of topological structures in the
misalignment scenario.

\begin{figure}
  \hfill
  \includegraphics[width=0.65\textwidth,bb=21 162 550 554]{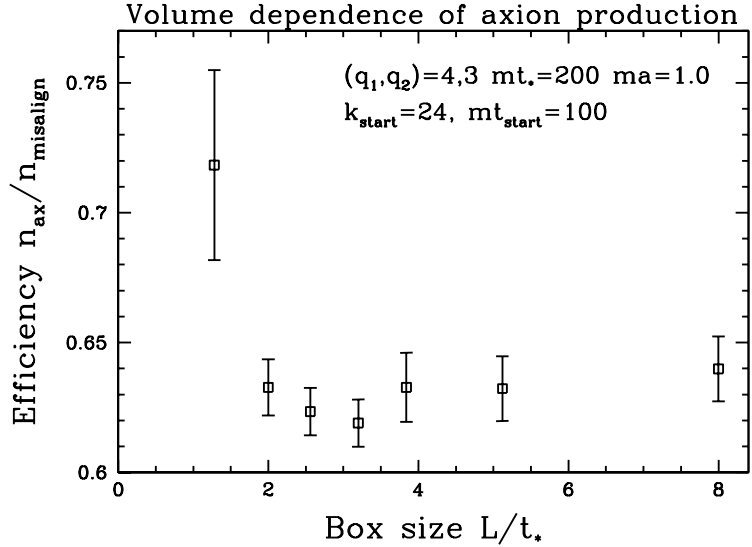}
  \hfill  $\phantom{.}$
  \caption{\label{fig:vols}
    Study of box volume dependence of the axion production rate.
    Except for the smallest volume, it appears the result shows
    extremely weak volume dependence.}
\end{figure}

The other test which proves to play almost no role in the final axion
density is the box volume.  To test out to very large and quite small
volumes, we used the rather small value of $mt_*=200$.  Keeping
everything else besides the volume unchanged, we find in Figure
\ref{fig:vols} that the volume has less than a $<2\%$ effect on the
axion abundance down to
a box length of $Lt_* = 2$.  Note that any box larger than $Lt_*=8$
should have essentially zero volume dependence, since the box
periodicity is invisible for $Lt>2$ and the axion number becomes an
adiabatic invariant by $t=4t_*$.

The smallest volume shown, $L=1.28t_*$, shows a larger generated axion
number, with larger statistical fluctuations (we used more simulations
for smaller volumes so the product of simulations and volumes is about
the same for each data point).  The reason is that, in a small
fraction of small-$L$ simulations, after all strings annihilate, there
remains a domain wall stretching across the whole box.  This
domain wall is metastable and lasts indefinitely, until it dominates
the axion number.  This is purely a small volume artifact; nothing of
the sort ever occurs for the larger volumes.

Because the volume dependence is so mild, it should be possible to
study the axion production in boxes down to $Lt_* =2$ or 3.
However, to be conservative, we have generally tried to keep
$Lt_* \geq 5$.  Recall that there cannot be lattice volume dependence
for $L > 2t$, and we need $t=4t_*$, so the box size dependence should
be exactly zero for $Lt_* > 8$.  The extremely weak box size
sensitivity allows us to relax this value somewhat.

\bibliographystyle{unsrt}
\bibliography{refs}

\end{document}